\documentclass[checkin,pre,showpacs]{revtex4}
\begin{document}        %
\draft
\title{Comments on "Evidence for Quantized Displacement in Macroscopic Nanomechanical Oscillators"} % Declares the title.
\author{A. Kwang-Hua CHU} %\thanks{Present Address : Department of Physics,
%Xinjiang University, 14, Road Shengli,Urumqi 830046, PR China.  }}  %address%\cite{PKU:1999}  %date
\affiliation{P.O. Box 30-15, Shanghai 200030, PR China \\ and \\
Department of Physics, Xinjiang University, 14, Road Shengli,
Urumqi 830046, PR China }
%%\renewcommand{\baselinestretch}{1}  %%\renewcommand{\baselinestretch}{1.5}
%\begin{document}           % End of preamble and beginning of text.
%%\renewcommand{\baselinestretch}{1}   %%\nopagebreak
\begin{abstract}
We make comments on Gaidarzhy {\it et al.}'s [{\it Phys. Rev. Lett.} 94, 030402
(2005)] letter.
%\noindent Keywords : Inflationary universe, brane inflation,
%string theory, anthropic principle .
%Magnus effects, vortex core, effective mass, critical
%Reynolds number, rarefied gases.
\end{abstract}
%%\end{titlepage}      %%\twocolumn     %%\nopagebreak   %\oddsidemargin=1mm
%\doublerulesep=6mm    %
%\baselineskip=6mm
\maketitle
\bibliographystyle{plain}
Quite recently Gaidarzhy {\it et al.} [1] have measured displacement in a
series of nanomechanical oscillators [2-3] with 1.49-GHz resonance
frequencies down to 110 mK temperatures. While
the low frequency mode at 21 MHz shows classical behavior
with expected drive dependence ($\propto B^2$), the 1.49 GHz
mode displays nonmonotonic dependence on driving force
at a temperature which corresponds to a thermal occupation
number $N_{th} \rightarrow 1$ (cf. Fig. 4 therein [1]).
They finally claimed that their experimental data indicate the
first observation of quantum displacement in macroscopic
nanomechanical oscillators. However, as they also argued that,
there may be other semiclassical
mechanisms which involve both nonlinearity and
quantum effects [4].\newline
The quantum mechanical harmonic oscillator [1-3] is a fundamental
example of textbook quantum mechanics.
Understanding the measurement of the quantum system itself requires identifying
the components of the quantum measurement process:
quantum system, measuring apparatus, and their
interaction [5]. Coupling to the environment, which introduces
decoherence and dissipation, must be included in the theoretical
analysis [6]. \newline
Based on the present author's experiences, we like to make some remarks about
the possible origin of their so-called {\it quantum displacement} in macroscopic
nanomechanical oscillators. Firstly, judged from the brief introduction of their
experimental procedure, their might be no explicit Casmir effects [7].
However, the zero-point vibrations at low temperature and the anharmonic
decay of the phonon [8] should be carefully taken into account as the authors already mentioned :
the intermediate jumps between these two (discrete) states
(in the range of $6.5-7$ Tesla in Fig. 4(b) [1]) could
represent transitions induced by thermal fluctuations, as
the temperature even at $N_{th}\rightarrow 1$ is high enough that the
thermal energy k$_B$T smears the gap hf in the energy levels. \newline
On the other hand, as noted by Gaidarzhy {\it et al.} in [1] : another possibility is
that the oscillator does not start
from the energy eigenbasis, as there is no {\it a priori} reason
for it to be in this preferred basis. The present author would like to point out,
the stochastic resonance [9-10], e.g., the two-state model [10],  might also
induce the results reported in Fig. 4 of [1]. Either the stochastic resonance
in a paramagnetically driven bistable buckling ribbon [11] or a magnetostochastic
resonance in thin films [12] have been reported previously. To be specific, considering
Fig. 4 (c)
in [1], as mentioned therein : two sweeps of the magnetic field, up (black) and down (blue), reproducing
the same transition, with occasional transitions to a possible
intermediate state, there also are possible weak stochastic resonances in the range of
$10-10.5$ Tesla (cf. Fig. 4 (c) in [1] together with Figs. 3 and 4 of Gammaitoni {\it et al.}
in [10]). \newline
In brief, detailed considerations of Casmir effects together with anharmonicity of thermal phonons
as well as stochastic resonances should be made to draw the final conclusion for the possible
observation of quantum displacement in [1].
%             %
%
%


\begin{thebibliography}{99} %\bibitem{PKU:1999}
\bibitem{Q:D} A. Gaidarzhy, G. Zolfagharkhani, R.L. Badzey, and P. Mohanty,
Phys. Rev. Lett. 94, 030402 (2005).
\bibitem{N:M} M. Blencowe, Phys. Rep. 395, 159 (2004).
\bibitem{Nano:Displ} R.H. Blick, F.W. Beil, E. H\"{o}hberger, A. Erbe, C. Weiss,
Nano-Electromechanical Systems: Displacement Detection
and the Mechanical Single Electron Shuttle, in {\it Lecture Notes in Physics}
(vol. 579, Springer, Berlin, 2001) pp. 215.
\bibitem{PRB:2004} V. Peano and M. Thorwart, Phys. Rev. B 70, 235401
(2004).
\bibitem{Q:DeCoher} W. Zurek, Rev. Mod. Phys. 75, 715 (2003).
\bibitem{Quantum:M} A. J. Leggett, J. Phys. Condens. Matter 14, R415 (2002).
A. J. Leggett, Phys. Scr. T102, 69 (2002).
P. A. M. Dirac, The Principles of Quantum Mechanics
(Clarendon , Oxford, 1981).
\bibitem{Casmir:2004} K.A. Milton, J. Phys. A : Math. Gen. 37, R209-R277 (2004).
\bibitem{Phonon:Anharm} M. Cardona, Ann. Phys. (Leipzig) 9, 865 (2000).
\bibitem{Stochas:Reson} T. Wellens, V. Shatokhin, and A. Buchleitner,
Rep. Prog. Phys. 67, 45-105 (2004).
\bibitem{S:R} L. Gammaitoni, P. H\"{a}nggi, P. Jung
and F. Marchesoni, Rev. Mod. Phys. 70, 223 (1998). Z. K.-H. Chu,
Eur. J. Phys.  21, L19-L20 (2000). {\it ibid}  23, L23-L24 (2002).
\bibitem{M:SR} M.L. Spano,  M. Wun-Fogle, and W.L. Ditto,  Phys.
Rev. A 46, R5253 (1992).
\bibitem{Film:SR} A.N. Grigorenko,  P.I. Nikitin, A.N. Slavin, and P.Y. Zhou,
 J. Appl. Phys. 76, 6335 (1994).
\end{thebibliography}
\end{document}